\documentclass[conference]{IEEEtran}
\IEEEoverridecommandlockouts
\usepackage{cite}
\usepackage{amsmath,amssymb,amsfonts}
\usepackage{algorithmic}
\usepackage{graphicx}

\usepackage{booktabs,makecell,tabularx}

\newcolumntype{C}[1]{>{\centering\arraybackslash}p{#1}}
\newcolumntype{L}{>{\raggedright\arraybackslash}X}

\usepackage{textcomp}
\usepackage{xcolor}
\def\BibTeX{{\rm B\kern-.05em{\sc i\kern-.025em b}\kern-.08em
    T\kern-.1667em\lower.7ex\hbox{E}\kern-.125emX}}
\begin{document}

\title{Decentralized Identifiers and \\ Self-sovereign Identity in 6G\\
}

\author{\IEEEauthorblockN{Sandro Rodriguez Garzon}
\IEEEauthorblockA{\textit{Service-centric Networking} \\
\textit{Technische Universit\"at Berlin / T-Labs}\\
Berlin, Germany \\
sandro.rodriguezgarzon@tu-berlin.de}
\and
\IEEEauthorblockN{Hakan Yildiz}
\IEEEauthorblockA{\textit{Service-centric Networking} \\
\textit{Technische Universit\"at Berlin / T-Labs}\\
Berlin, Germany \\
hakan.yildiz@tu-berlin.de}
\and
\IEEEauthorblockN{Axel K\"upper}
\IEEEauthorblockA{\textit{Service-centric Networking} \\
\textit{Technische Universit\"at Berlin / T-Labs}\\
Berlin, Germany \\
axel.kuepper@tu-berlin.de}
}

\newcommand{\ts}{\textsuperscript}

\maketitle

\begin{abstract}
A key challenge for mobile network operators in 6G is to bring together and orchestrate a variety of new emerging players of today's mobile ecosystems in order to provide economically viable and seamless mobile connectivity in form of a multi-stakeholder service. With each new player, be it a cloud, edge or hardware provider, the need for interfaces with secure authentication and authorization mechanisms increases, as does the complexity and operational costs of the public key infrastructures required for the identity and key management. While today's centralized public key infrastructures have proven to be technically feasible in confined and trusted spaces, they do not provide the required security for access control once centralized identity providers must be avoided because of limited cross-domain interoperability, national data protection legislation, or geopolitical-strategic reasons. Recent decentralized identity management concepts, such as the W3C recommendation of Decentralized Identifiers, provide a secure, tamper-proof, and cross-domain identity management alternative for future multi-stakeholder 6G networks without relying on centralized identity provider or certification authorities. This article introduces the concept of Decentralized Identifiers together with the principles of Self-sovereign Identity and discusses opportunities and potential benefits of their application and usage for cross-domain and privacy-preserving identity and key management in 6G networks.
\end{abstract}


\section{Introduction}
Over the past decades, public landline mobile networks (PLMN) have experienced a major structural transformation. They have evolved from closed monolithic hardware and software systems, owned, and managed by dedicated mobile network operators (MNOs), to open, distributed and interconnected systems with a wide variety of actors and stakeholders involved. The liberalization of PLMN ecosystems and the network disaggregation will continue in the future, and new players, such as spectrum brokers, cable, satellite, IoT, or over-the-top (OTT) service providers will take on important roles in the provision of mobile connectivity and related services within converged networks. Consequently, the key role of MNOs needs to be redefined from a sole network operator to a connectivity provider that is responsible to orchestrate the interaction of diverse and sometimes competitive players within the mobile connectivity market. This makes 6G connectivity a multi-domain service that relies on resources distributed across multiple trust domains \cite{b1}. So, the more actors are engaged in the operation of a disaggregated and open PLMN, the more cross-domain interfaces are required, and the less do the access and core networks correspond to their original form of closed trusted environments. As a result, in 6G, PLMNs must consequently be considered as zero-trust architectures (ZTAs). But since trustworthiness is one the key design principles of 6G \cite{b2}, trust must be established between the parties and their network components involved through unified access management. 

However, the long specification periods and the incremental opening of PLMN ecosystems have led to a wide variety of authentication or authorization procedures being used in PLMNs today. While mobile subscribers still authenticate in 5G using symmetric cryptography, in roaming, the mutual authentication between MNOs can be carried out with TLS, based on asymmetric cryptography with pre-shared public keys. To apply asymmetric cryptography for authentication purposes throughout the PLMN, public key infrastructures (PKIs) need to be put in place, be operated at each MNO, and be securely interconnected. Besides the additional administrative burden and operational costs there remains a lack of commonly trusted certification authorities (CAs) on a global level. The reasons are manifold, ranging from geopolitical differences to the avoidance of a single point of failure or attack. In 5G, this leads MNOs, as recommended by the GSMA \cite{b3}, to operate at least one root CA, e.g., for roaming purposes. Nonetheless, the lack of commonly trusted CAs renders the federated identity management approach with interconnected centralized PKIs for 6G complex and impractical. 

\begin{table}[!htb]
	\small
	\setlength{\tabcolsep}{3pt}
	
	\begin{tabularx}{\linewidth}{ l l }
		\toprule
		\thead{Acronym}  & \thead{Definition} \\
		\midrule
		3GPP & 3rd Generation Partnership Project \\
		5G	& 5th Generation Technology Standard \\
		6G 	& 6th Generation Technology Standard \\
		CA	& Certification Authority \\
		CPE	& Customer-premises Equipment\\
		DID	& Decentralized Identifier \\
		DL	& Distributed Ledger \\
		DLT	& Distributed Ledger Technology \\
		GDPR &	General Data Protection Regulation \\
		GSMA &	Global System for Mobile Communications Association \\
		IDM	& Identity Management \\
		IoT	& Internet of Things \\
		IPX	& Internetwork Packet Exchange \\
		IS	& Identity Subject \\
		MNO	& Mobile Network Operator \\
		NF	& Network Function \\
		NFV	& Network Function Virtualization \\
		NRF	& Network Repository Function \\
		O-RAN	& Open Radio Access Network \\
		OTT	& Over-the-Top \\
		PKI	& Public Key Infrastructure \\
		PLMN &	Public Landline Mobile Network \\
		PRINS &	Protocol for N32 interconnect security \\
		RAN	& Radio Access Network \\
		SBA	& Service Based Architecture \\
		SIM	& Subscriber Identity Module \\
		SSI	& Self-Sovereign Identity \\
		TLS	& Transport Layer Security \\
		UDM	& Unified Data Management \\
		VC	& Verifiable Credential \\
		W3C	& World Wide Web Consortium \\
		ZTA	& Zero-trust Architecture \\
		\bottomrule
	\end{tabularx}
	\label{table_acronyms}
	\caption{Abbreviations}
\end{table}

Meanwhile, subscribers started to pay attention on how their personal data is handled by service providers, resulting in new regional privacy regulations to be applied, such as the GDPR in Europe. With more parties involved in the provision of connectivity and related services, compliance with the new privacy and transparency-enforcing regulations becomes increasingly challenging for MNOs because personal data is centrally hold and processed by MNOs and its partners. Hence, new methodologies to treat identity information in a privacy-preserving manner need to be developed and applied in multi-stakeholder 6G networks to increase transparency for all actors, including the subscribers, and to comply with the latest legal frameworks for the collection and processing of personal information \cite{b4}. 

\begin{figure*}[!htp]
	\centerline{\includegraphics[scale=0.37]{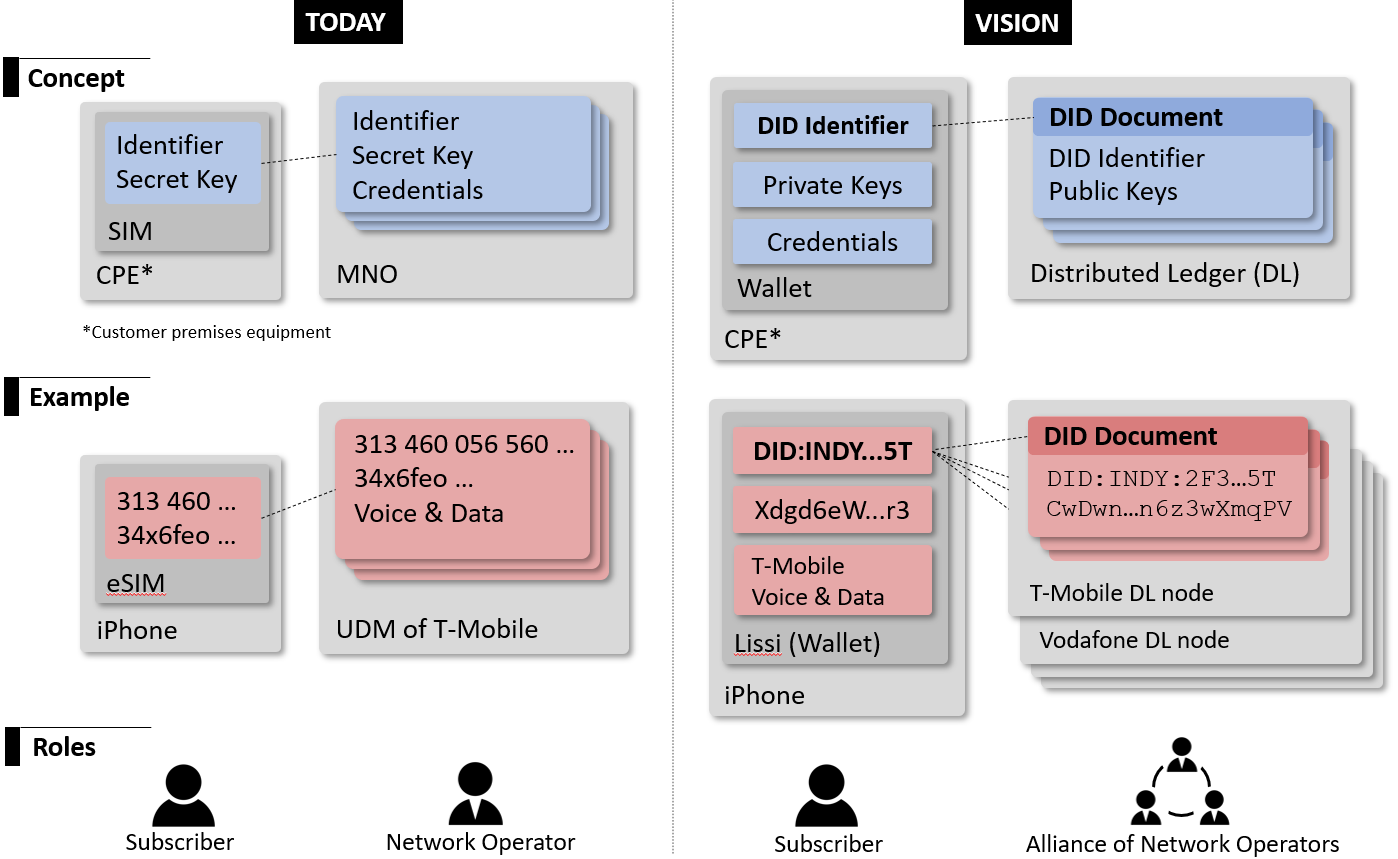}}
	\caption{Centralized (today) vs. decentralized IDM with DIDs in 6G (vision). [iPhone is a registered trademark of Apple Inc.; T-Mobile is a registered trademark of Deutsche Telekom AG; Vodafone is a registered trademark of Vodafone Group PLC.; Lissi: https://lissi.id]}
	\label{fig1}
\end{figure*} 

Accordingly, due to the heterogeneity of today's identity management (IDM) systems in 4G/5G, the lack of globally trusted CAs and the ever-increasing demand of identity subjects (IS) for personal data protection, a fundamentally different and unified approach to IDM must be adopted in 6G. The digital identities of 6G must meet the following fundamental requirements to be suitable for future-proof IDM in multi-stakeholder 6G networks:   

\begin{itemize}
	\item {\bf{Universality}}: They do not only represent subscribers, but any IS that is involved in the usage and provision of mobile connectivity and related services, including but not limited to IoT devices, MNOs, virtual mobile network, cloud or IoT operator, manufacturer, or even network functions (NF).
	\item {\bf{Privacy-preserving}}: They neither reveal the real identity of an IS nor consist of any personal data related to the real identity of an IS.  
	\item {\bf{Interoperability}}: They are all unique, machine-readable, of standardized format, and commonly accessible.  
	\item {\bf{Verifiability}}: They are cryptographically verifiable to enable the referred IS to proof ownership over the digital identity and others to proof the validity of personal data shared by the referred IS. 
	\item {\bf{Extensibility}}: They are extensible with respect to the supported verification methods due to the diversity of potential application areas.
	\item {\bf{Independence}}: They can be created, edited, and used for verification purposes independently of corporates, organizations, and governments, including identity provider and CAs.
\end{itemize}

\noindent However, to achieve independence, IDM in 6G needs to be organized in an entirely decentralized fashion via a peer-to-peer network, where each peer is owned and operated by a single stakeholder but the IDM executing atop the peer-to-peer network as a decentralized application is being governed equally by all involved stakeholders. 

With the recommendation of Decentralized Identifiers (DID), the W3C has recently introduced a promising method to manage identities in a decentralized manner without centralized identity providers and CAs via, e.g., a distributed ledger (DL) \cite{b5}. The decentralization aspect of DIDs bears the inherent potential to unify access management procedures in 6G across-domains by using a DL among all involved actors of a PLMN ecosystem as the single source of truth for non-privacy-sensitive access management data. A DID infrastructure makes it also possible to store access authorizations according to the principles of Self-Sovereign Identity (SSI) \cite{b6} privacy-preserving as verifiable credentials not centrally, e.g., within the UDM, but exclusively at the authorized party, e.g., within the CPE. Applying DIDs in 6G to decentralize IDM will not only lower the risk of having a single point of failure or attack due to the removal of CAs but they will also enable MNOs to adhere to privacy-by-design principles \cite{b7} with SSI. In addition, DID and SSI have the potential to reduce the operational overhead and complexity, and to increase the security and reliability of cross-domain IDM and, consequently, to foster the harmonization of authentication and authorization mechanisms in future highly heterogeneous 6G networks. 

This article introduces the DID concept and SSI paradigm in the context of 6G, describes how they meet the requirements for IDM in 6G and discusses potential opportunities in form of concrete use cases on the access, management, and application planes. Table \ref{table_acronyms} presents the list of abbreviations used in this article. 

\section{Decentralized Identifiers}


A DID is an unique identifier, consists of an alphanumeric string, refers to an IS (here denoted as a {\bf{DID subject}}), and resolves to a {\bf{DID document}}. A DID subject can be a person or any type of entity, e.g., a subscriber, an MNO or a NF. A DID document comprises among other information, verification material, e.g., a public key of the DID subject for authentication, authorization, and other purposes. Hence, a DID binds a DID subject to its DID document just like a X.509 certificate binds an identity to a public key. Neither the DID nor the DID document should reveal or contain any personal data. The fundamental idea of the DID concept is to manage and share non-privacy-sensitive verification material in the form of DID documents among interacting stakeholders via a DL while the associated privacy-sensitive personal data of a DID subject is kept locally, off the DL, ideally at the DID subject's respectively stakeholder's premises. Hence, the DL is used to persist and share identity-related data in a fully synchronized and tamper-proof manner among all actors of a multi-stakeholder network. Already persisted data can't be removed, and new data only be added. Since the data is continuously replicated across all nodes of the multi-stakeholder network, all stakeholders have access to it. 

\begin{figure*}[tp]
	\centerline{\includegraphics[scale=0.40]{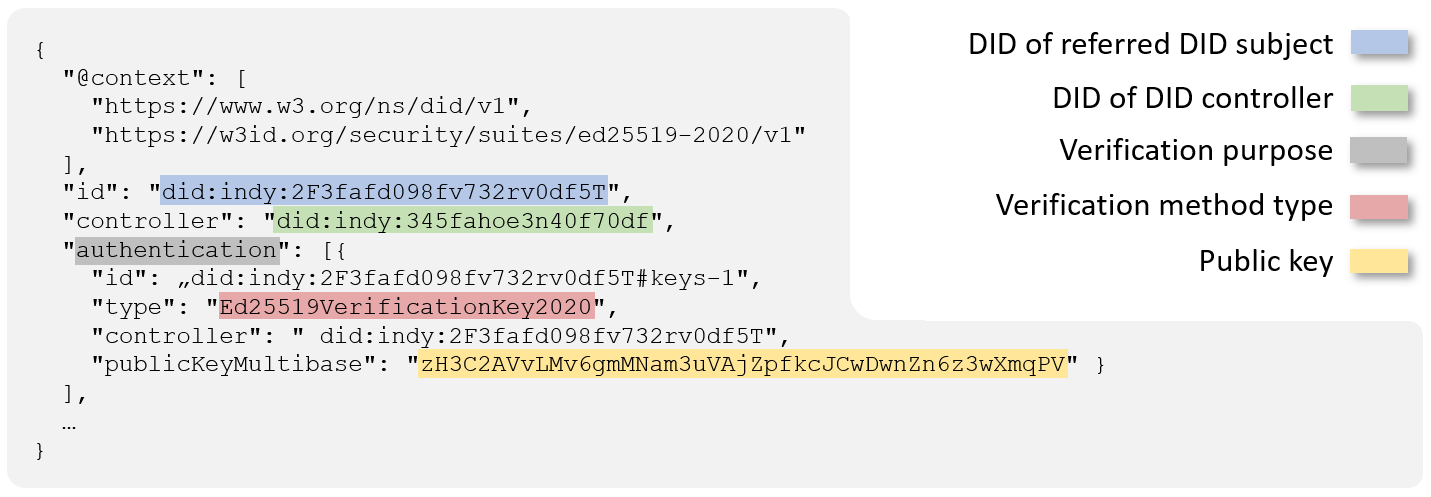}}
	\caption{Exemplary DID document}
	\label{fig2}
\end{figure*}

Due to the inherent characteristics of DL technology (DLT), there is neither a dedicated identity provider nor a CA needed to enable the verifiability of the DID documents. Although DID documents are accessible by all nodes of the multi-stakeholder network, they can only be updated (data added to make old versions obsolete) either by the DID subject via its locally kept private key or by a deputy explicitly appointed by the DID subject, called {\bf{DID controller}}. With the securely stored private key and the generally accessible public key in the DID document, a DID subject can prove ownership of its DID document to other DID subjects. In other words, A DID subject can authenticate itself towards other DID subjects. The DID concept is also extensible, e.g., with respect to the supported verification methods. A schematic overview of the DID concept is given in Figure \ref{fig1} and an exemplary DID document in Figure \ref{fig2}.

The above-mentioned requirements for IDM in 6G networks regarding universality, privacy-preserving, interoperability, extensibility, and independence are fully met by W3C DIDs. By enabling a DID subject to identify itself to others as being the owner of the digital identity, DIDs also support one of the two essential aspects of verifiability. Since the DID concept specifies all functions of a decentralized PKI, it disrupts its centralized counterparts and makes certificates for authentication purposes, e.g., X.509, with centralized CAs entirely obsolete. Instead, it makes use of DLT to enforce trust algorithmically between all stakeholders governing a common DL for IDM purposes. 

\subsection{DIDs on the Management and Access Planes}

DIDs can be applied to support the key exchange procedures between MNOs and between MNOs and IPX providers for roaming purposes. If MNOs and IPX providers are represented by DIDs on a common DL, then the public keys for the mutual authentication can be placed in the associated DID documents. In case public keys need to be rotated for security reasons, DID documents will be updated and synchronized automatically between all nodes the multi-stakeholder network. However, the DIDs need to be exchanged prior to the initial authentication. Afterwards, each party can proof ownership of a DID towards another party by means of the securely owned private keys. The DID specification allows optionally to define service end points in the DID documents. A service end point is meant to be the network address at which a DID subject can be contacted at. Changes of gates for roaming signaling can then be securely communicated via changes of the DID document's service end points.

A apparent appliance of DIDs in 6G is the DID-based subscriber authentication. Xu et al. introduce a new CA-less subscriber authentication scheme \cite{b8}. Although not relying on W3C DIDs, Xu et al. could at least show that subscriber authentication is technically feasible without CAs by decentralizing subscriber IDM. Haddad et al. follows a similar approach and introduces a new authentication and key agreement protocol for 5G that makes use of blockchain-enabled IDM \cite{b9}. Raju et al. could even proof that the signaling traffic related to network access can be reduced by up to 40\% in cognitive cellular networks if IDM is organized in a decentralized fashion \cite{b10}. Gnomon \cite{b13} and the DIAM-IoT framework \cite{b15} both apply the latest W3C DID recommendation within their DLT-based IDM solutions for the IoT device registration process 

\begin{figure}[tp]
	\centerline{\includegraphics[scale=0.32]{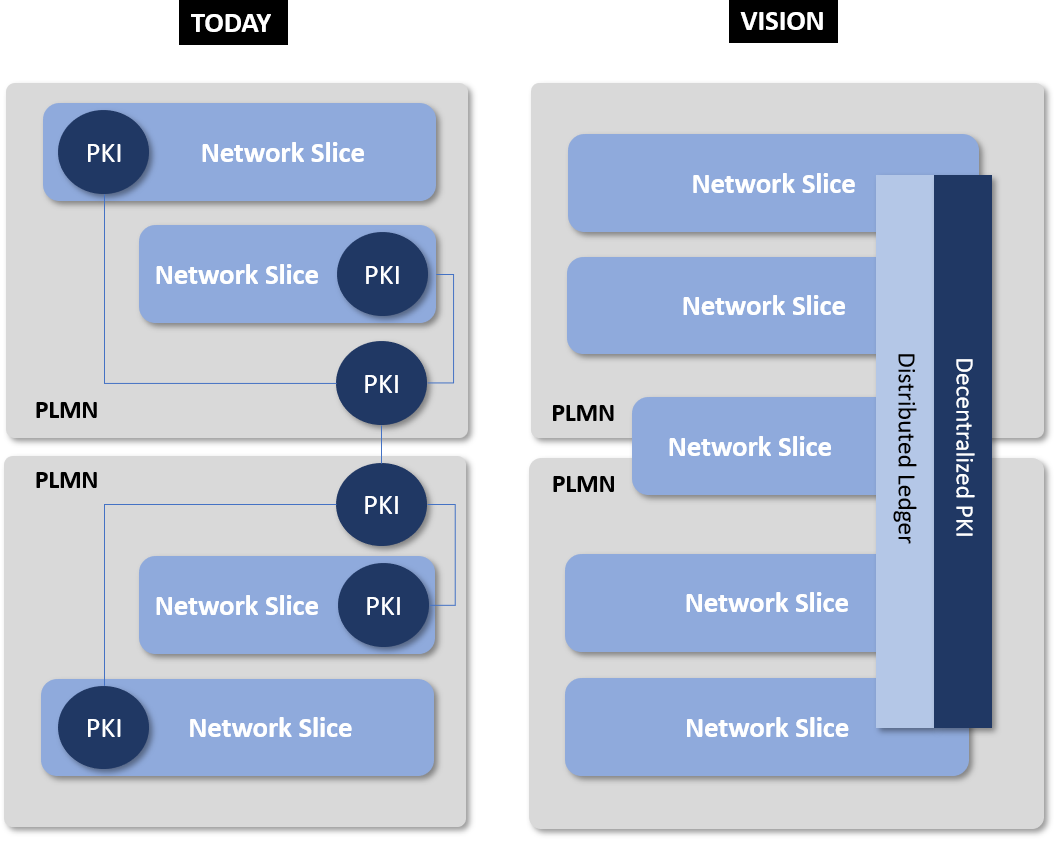}}
	\caption{Cross-slice and cross-PLMN decentralized PKI with DIDs}
	\label{fig3}
\end{figure}

Since network slicing leads naturally to a core network that is shared by multiple actors like virtual network operators, authentication between NFs becomes mandatory. Today, the mutual authentication is conducted with TLS and CA-issued certificates. In roaming, MNOs are profiting from their collaboration, but virtual network operators are competing on the same territory. To agree on a commonly trusted CA, even if it is the MNO itself, becomes challenging. At least for network slices spanning different trust domains, commonly trusted CAs are required if centralized PKI technology is applied. Following the decentralized principles of DIDs, on the other hand, can enable a CA-less key management across all operators of virtual networks or network slices in a future SBA. Figure \ref{fig3} illustrates the concept of having a cross-slice and cross-PLMN decentralized PKI with DIDs.  

The O-RAN alliance considered RAN to be built as a ZTA with TLS and X.509 certificates for the mutual authentication between components of a disaggregated base station, comprising cloud native and physical NFs. O-RANs will be geographically distributed which makes them predestined to operate the required key management in a decentralized fashion with DIDs at the edge of the network instead of a centralized PKI within the core network.

The loose decoupling between software components and physical equipment as proclaimed in O-RAN and already introduced as NFV in 4G and 5G core networks adds additional demarcation lines between hardware, cloud, virtualization, and NF providers. Mutual authentication is therefore not exclusively needed on the functional layer, but up the system chain starting from the physical equipment, over the hypervisor management and the orchestration layers, till the software-defined networking and virtual NF layers. Since NFV developments are still in an early stage, decentralized key management can be considered as a viable solution from the very beginning.  

\subsection{DIDs on the Application Plane}

The spectrum of opportunities for DID-based authentication is not limited to intra-PLMN or inter-PLMN communication on the control plane, but also encompasses any authentication on the user plane between subscribers and OTT services. While originally, service users had to register with an OTT service exclusively, and thus had to create a digital identity explicitly for this OTT service, nowadays, 3\ts{rd} parties take on the role of general-purpose identity providers, with Google and Facebook being among the most popular incarnations. These centralized identity providers act as commonly trusted 3\ts{rd} parties and they do not only provide the identifier, but also manage privacy-sensitive personal data and access to it by other OTT services. With the introduction of the GDPR and the ever-increasing need among service users for more transparency and security regarding digital identity data, the centralized approach via a few international players seems outdated. Not only because service users outsource personal data to external and non-national identity providers - in conflict with the latest privacy-by-design principles -, but also because this makes them dependent on the provider's security concepts, business practices, privacy regulations, and the legislation of foreign countries.  

Fortunately, the concept of DIDs offers a security- and privacy-compliant and independent solution for IDM at the user plane that is safe from enterprise and state intervention. It also enables new non-OTT actors to participate in the value chain around IDM for OTT services. For example, national MNOs can act as DID controllers for their subscribers in the sense of national trust anchors. They can also support resource-constrained terminals in DID-based authentication towards OTT services by providing a network-centric and secure access to the DL. The latter becomes necessary when terminals do not have sufficient technical resources to participate directly in the operation of the DL but must obtain the public keys of the OTT services (anchored in the related DID documents) for authentication purposes. Hereby, MNOs can participate in the operation and governance of the DL and provide exclusive and trusted access to it for their subscribers.  

In the IoT domain, IDM became critical for the interoperability of IoT devices from different manufactures and operators. For autonomous driving or flying with drones, mobile as well as infrastructural IoT devices need to mutually authenticate each other in an ad-hoc fashion to exchange environment information in a secure and trustful manner. Taking a decentralized approach to IDM with DIDs in IoT ecosystems would foster the interoperability through the harmonization of ad-hoc authentication mechanisms and enables MNOs to play a pivotal role in IDM for IoT devices through the operation of the required DL and provision of secure access to it in 6G networks. Fedrecheski et al. give an outlook of applying DID-based IDM to the domain of IoT devices by comparing it with the current certificate-based approaches \cite{b11}.  

\begin{figure*}[!htp]
	\centerline{\includegraphics[scale=0.46]{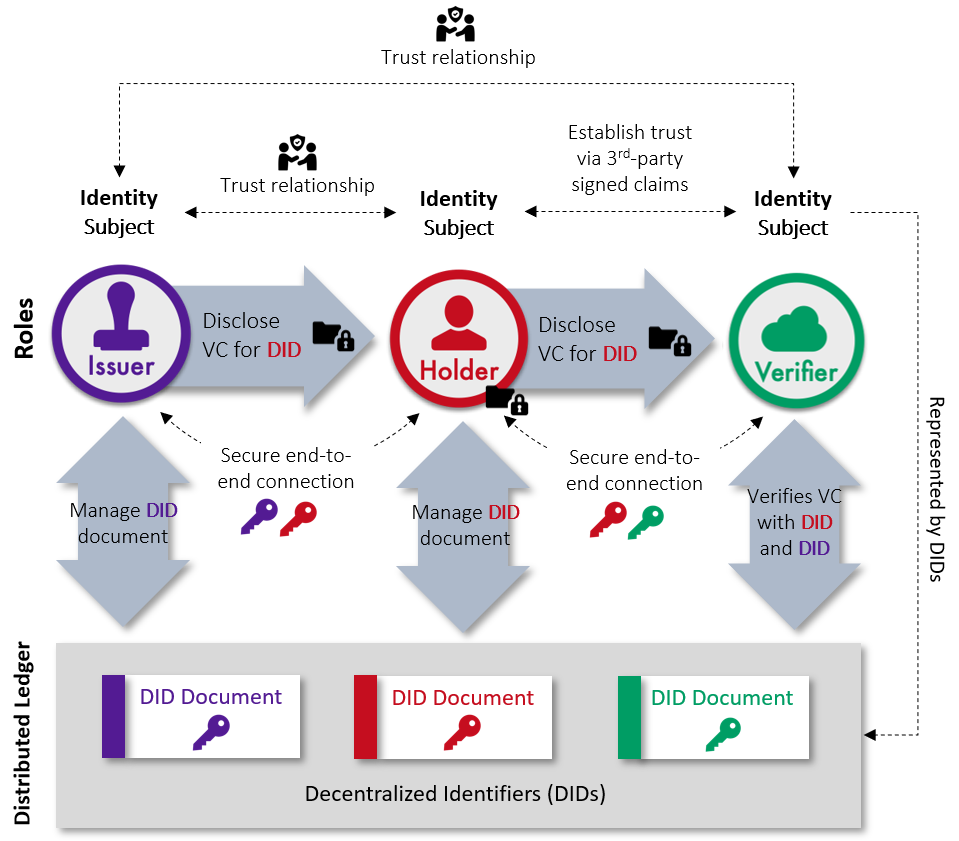}}
	\caption{Self-sovereign Identity methodology with DIDs}
	\label{fig4}
\end{figure*} 

\section{Self-sovereign Identity}

Self-sovereignty over any information describing an entity is the principal objective behind the latest SSI paradigm. All digital information which are related to an identified or identifiable subject, e.g., the social security number, should ideally be held and managed solely by the IS. With SSI, the decision whether at all, to whom and to what extent personal data ({\bf{credentials}}) are disclosed lies solely with the IS. To establish a trust relationship between two IS's that have not yet interacted with each other, it is beneficial for IS's to exchange credentials that has been approved by a 3\ts{rd} party that are equally trusted by both IS's. So, for the purpose of trust building with SSI, an IS (role: {\bf{holder}}) can disclose to another IS (role: {\bf{verifier}}) its very own credentials in the form of {\bf{verifiable credentials}} (VC) \cite{b12} that has been previously approved, digitally signed and handed over to the holder by a 3\ts{rd} IS (role: {\bf{issuer}}). For example, a potential customer (holder) discloses its social security number, as issued by governmental authorities (issuer) as a VC, to an MNO (verifier) for setting up a mobile phone plan. The trust triangle in SSI is only viable if the issuer's approval of the disclosed claim is cryptographically verifiable by the actor consuming the credentials (the verifier). SSI thus not only proclaims that privacy-sensitive credentials should be hold and managed by solely the IS, but that the credentials could also be approved by a trustful 3\ts{rd} party and be verified by others. This makes SSI a new fundamental concept to enforce privacy preservation and to enable trust building between actors of different trust domains. 

SSI can be implemented in a wide variety of ways, e.g., with certificates and centralized CAs. However, the conceptual separation between the commonly accessible DID documents on a DL and the privacy-sensitive credentials, as remaining with the ISs, accurately reflects the technical requirements for SSI. Hence, DIDs provide naturally a technical mechanism to implement a decentralized IDM system based on the principles of SSI, without relying on a single identity provider or CA. With DIDs only, two ISs can establish a secure point-to-point connection with mutual authentication. With SSI, they can in addition share digitally signed VCs off the DL via the established connection and verify the VCs using the public keys in the commonly accessible DID documents. DID documents thus do not only contain public keys to authenticate an IS, as described above, but also to verify the authenticity of VCs. VCs thus cover the outstanding aspect of the verifiability requirement for digital identities in 6G, namely the verifiability of personal data by 3\ts{rd} parties. Figure \ref{fig4} illustrates the SSI methodology if DIDs are used as the technical foundation.

\subsection{SSI in the Access and Management Planes}
MNOs, their technical partners, and the subscribers will benefit from the compliance with SSI principles in 6G in many ways. For MNOs and their partners, SSI can bring efficiency improvements, architectural simplifications, and security enhancements while subscribers can experience new and privacy-preserving features in 6G.    

With SSI, the change of terminals would be simplified if access permissions are not bound to hardware-based SIM cards, but to the personal identity of the subscriber. Given both MNOs and subscribers are represented by DIDs, an MNO will be able to issue access permissions in the form of VCs directly to the subscriber while the necessary public keys for verification purposes are stored in a tamper-proof manner in the MNOs DID documents. If the subscriber requests connectivity in a roaming situation, the subscriber (holder) hands over the VCs to the visited MNO (verifier) which in turn verifies its validity by means of the home MNO's public keys in the DID documents. Hence, a change of a terminal device will not lead to a transfer of a SIM card, but to a transfer of the access permission VCs. The latter step is lesser critical because the VCs are anyhow bound to a dedicated DID. Malicious subscribers can't make use of the VCs unless they act on behalf of the subscriber in the role of a DID controller or got access to the private key of the subscriber's DID. Since a DL enforces continuous synchronization  of the data across all nodes, a potential revocation of access permissions by the home MNO would almost instantly be taken notice of by any visited MNO through updated entries in the home MNO's DID documents.  

\begin{figure*}[!htp]
	\centerline{\includegraphics[scale=0.44]{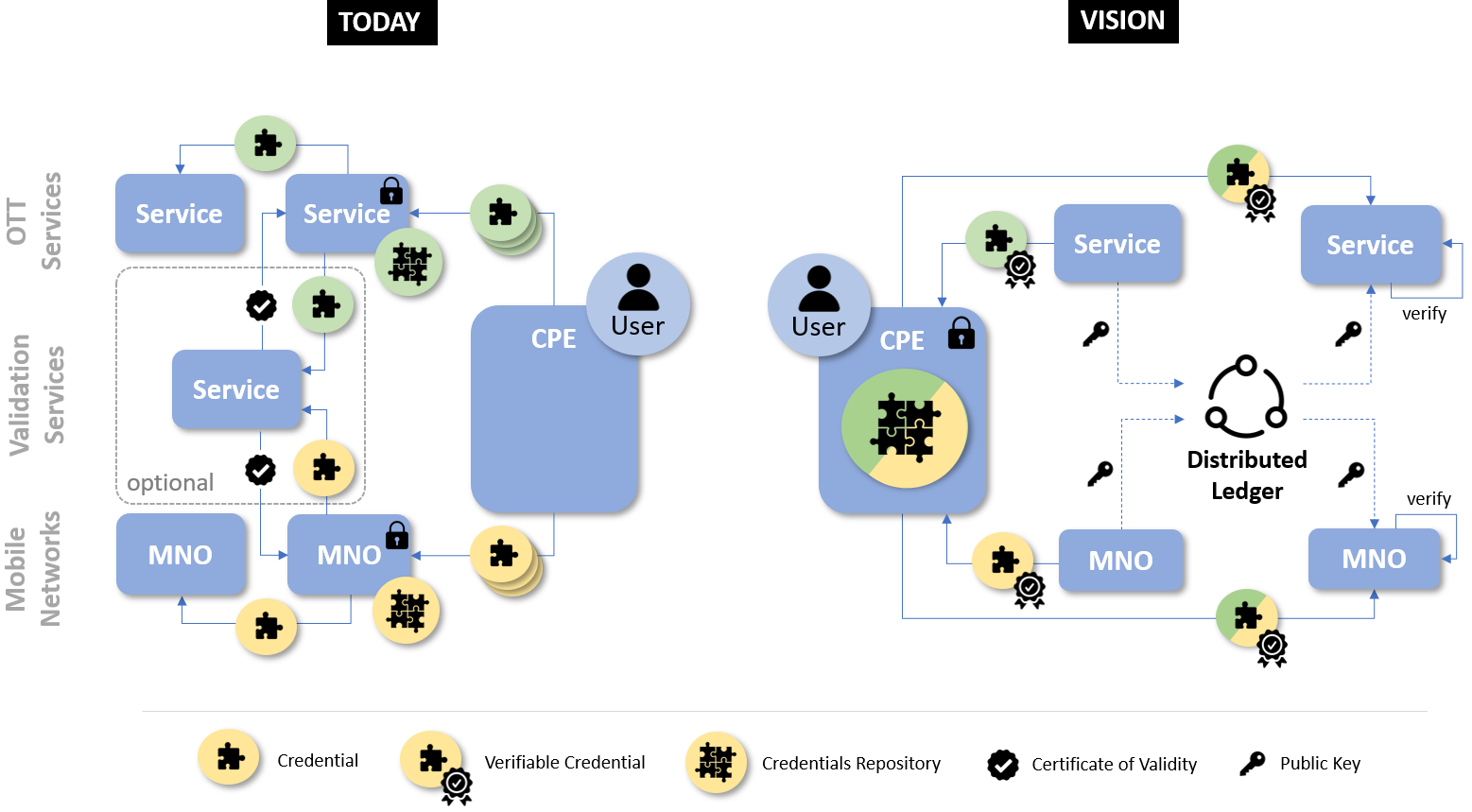}}
	\caption{Decoupled centralized IDM (today) vs. unified decentralized IDM according to SSI principles (vision)}
	\label{fig5}
\end{figure*}

While the subscriber experiences more flexibility with respect to personal mobility, interpreting access permissions as VCs makes it also possible from the visited MNO's perspective to authorize network access to guests without querying the home MNO [9]. Everything needed for a visited MNO to securely authorize a guest are the VCs as provided by the guest and an access to the DL to query the home MNO's DID document. Moreover, even in a non-roaming scenario, a home MNO will benefit from access permissions as VCs. The subscriber authentication and authorization could be executed in a decentralized fashion at the edge only, without involving the core network. Since several thousand of IoT devices are expected to register at a single base station in 6G, this would significantly reduce the load on the backhaul links. Salleras et al. make even use of the principles of decentralization to authorize network slice users towards the network slice without letting the users disclose any privacy-sensitive credentials beyond the access permissions \cite{b14}.   

A network authorization, whether a visited MNO is permitted to serve guests from a specific home MNO, can also be encoded as VCs. In this case, a home MNO acts in the role of the issuer and hands over permissions to serve their customers in the form VCs to a partnering MNO (holder). A subscriber (verifier) of the home MNO can then always check whether the VCs provided by a partnering MNO during connection establishment are still valid by querying the necessary DID documents. Network authorizations in form of VCs can even simplify the process of giving partnering non-3GPP access networks the opportunity to serve guests of a home MNO without requiring a direct link to a 3GPP network, such as the home MNO's PLMN. 

Despite the SSI paradigm being initially introduced to preserve privacy with respect to digital identity-related data for humans, it can also be applied in the management plane of future PLMNs to share authorizations in the form of VCs. In 5G roaming, IPX providers can be actively involved in the signaling between a home and a visited MNO. With the PRINS, MNOs can grant partnering IPXs rights to alter, on their behalf, the information elements of signaling messages between MNOs and to make the IPX-induced changes verifiable at the message-receiving MNO. PRINS requires complex PKI infrastructures to be present at each MNO and IPX provider and the public keys to be pre-shared among the involved actors. With SSI, authorizations to alter information elements can be issued by the MNO to its partnering IPX providers in the form of VCs while the message-receiving MNO can verify the VCs via the DL. This replaces the complex setup of interacting PKIs of different MNOs and IPX providers with a common decentralized PKI in form of a single cross-domain DL.  

Another opportunity to apply SSI principles in 6G in the management plane concerns the NF-to-NF authorization via OAuth 2.0 in the present SBA. Currently, the NRF acts as a trust anchor and provisions authorizations to access resources of a NF to requesting NFs as JSON Web Tokens. A consuming NF attaches the JSON Web Token to the request towards the producing NF, which in turn verifies the validity of the JSON Web Token through a CA. Likewise, at least a single PKI is needed for these purposes within the SBA and multiple interacting PKIs of different MNOs for future cross-SBA network slices. Instead, JSON Web Tokens can be replaced by VCs issued by the NRF, hold by the consuming NF, and be verified by the producing NF in case of a service request. The latter example demonstrates that the IS does not necessarily need to be a single human or enterprise to generate benefits in terms of unified authentication/authorization procedures and the elimination of a single point of failure but also a software component in form of an NF or in future multiple-access edge cloud scenarios in the form of 3\ts{rd} party edge cloud application instances.

\subsection{SSI in the Application Plane}

A subscriber's benefit of SSI in 6G stems from the possibility to use MNO-collected and -provided context information securely and privacy-preserving with OTT services. MNOs can, for example, issue a subscriber (holder) a VC stating the current geographical location of the subscriber as determined and approved by the MNO. The subscriber is then free to disclose its verifiable location as a VC with any OTT service that requires an MNO-approved location to provide its services. The verification is done by the OTT service via the MNO's DID documents. The MNO's role as an issuer for trusted context information opens even new potential revenue streams for MNOs since the context information is already present in PLMNs but only utilized for providing connectivity. Vice versa, MNOs can also benefit from VCs that are issued by OTT services to the subscriber. Taking the example of above, an MNO in the role of a verifier can simplify and thus reduce costs associated with the know-your-customer process during the onboarding of a new customer by verifying the social security number (in form of a VC) as issued by governmental authorities. A costly call to a 3\ts{rd} party service to verify the credentials of the new customer is no longer necessary. The potential MNO roles within a unified IDM with DIDs and SSI on the application plane are illustrated in Figure \ref{fig5}.   

\section{Conclusion and Future Work}

The increased complexity of interconnected PKIs, the lack of commonly trusted CAs and the ever-growing demand for privacy protection led to new fundamental requirements for future-proof IDM in 6G networks. The concept of DIDs provides thereby all required conceptual ingredients to design and implement a common, cross-domain, and tamper-proof IDM system for unified and secure intra- and inter-PLMN authentication purposes, without centralized identity providers or CAs. With DID and SSI, 6G networks would even be able to implement privacy-preserving and trust building mechanisms in the access, management, and application planes in which credentials of humans or things are signed by commonly trusted issuers, hold by the IS's according to privacy-by-design principles and still be independently verifiable by 3\ts{rd} parties. The combination of DID and SSI bears therefore not only the potential to make today's interconnected and centralized PKIs in 4G/5G obsolete but it, in addition, opens up the opportunity for MNOs to provision trustful and privacy-preserving mobile connectivity and services in future multi-party zero-trust environments. 

This article does by no means present a complete list of applications and benefits of DID and SSI for 6G but gives a first glance on the impact decentralized IDM will have on the aspects of trust, security, and privacy at all levels within future 6G networks. Other potential applications on the management plane are, e.g., the DID and SSI-based authentication and authorization of application instances within and across network slices or directly in the mobile edge cloud, and, in the application plane, the mutual DID-based authentication and SSI-based exchange of VCs between subscribers. However, further research is needed and must address open research questions with respect to the technical implementation and architectural integration of DLT in PLMN ecosystems. Further investigations must also cover governance and standardization aspects of operating a DL across competing actors in the PLMN ecosystems and OTT services, and must address the operational overhead induced by DLT, the ability of DLT to scale to billions of subscribers and the expected speed with which DIDs can be created, altered, and synchronized reliably among all nodes of a decentralized IDM system.

\vspace{12pt}

\section*{Biographies}

\noindent SANDRO RODRIGUEZ GARZON is a senior researcher at the Technische Universit\"at Berlin (TU Berlin) and T-Labs. He received the Dipl.-Ing. degree in computer engineering from the TU Berlin in 2005 and the doctorate from the TU Berlin in 2013. From 2006 to 2007, he worked as a technical consultant for the sd\&m AG (now Capgemini sd\&m AG) and has participated in IT projects for OEMs such as Volkswagen and Audi. From 2007 to 2013, he worked as a research associate with the Daimler Center for Automotive IT Innovations with a three-month research guest visit at Mercedes-Benz Research and Development North America, Palo Alto, CA, USA. Since 2013, he has been working with the T-Labs within the group of Service-centric Networking at the TU Berlin.

HAKAN YILDIZ is a research associate at the Technische Universit\"at Berlin (TU Berlin) and T-Labs. He received his Dipl.-Wirt. -Ing. degree from the Karlsruhe Institute of Technology in 2011. Coming from an industrial engineering background, he has worked for various corporations, including Robert Bosch GmbH and HERE Technologies GmbH, in product and project management roles. Hakan joined the Service-centric Networking chair of the TU Berlin in 2020 to specialize in DLT. Since then, he has been researching the interoperability of SSI solutions while working on the research Project IDunion, funded by the German Ministry of Economics and Energy (BMWi).

AXEL K\"UPPER is a professor for Service-centric Networking at the Technische Universit\"at Berlin (TU Berlin) and T-Labs, with the latter being a public-private partnership of Deutsche Telekom AG and TU Berlin. Before, he was an assistant professor at Ludwig-Maximilians-Universit\"at M\"unchen, where he received his postdoctoral lecture qualification (habilitation). He graduated in computer science from RWTH Aachen University, where he also received his doctorate. Axel is an advocate of decentralized systems and applications. The focus of his work lies on DLT, token economy, SSI, emerging blockchain-based applications, blockchain analytics, and decentralized online social networks.  


\begin{thebibliography}{00}
\bibitem{b1} E. Bertin, N. Crespi, and T. Magedanz, Eds., {\it{Shaping Future 6G Networks: Needs, Impacts, and Technologies}}: IEEE Press Wiley, 2021.
\bibitem{b2} W. Tong and P. Zhu, {\it{6G: The next horizon: From connected people and things to connected intelligence}}. Cambridge: Cambridge University Press, 2021. 
\bibitem{b3} GSM Association, {\it{Key Management for 4G and 5G inter-PLMN Security}}, 2022. [Online]. Available: https://www.gsma.com/security/wp-content/uploads/2022/05/FS.34-v4.0.pdf (accessed: July 19 2022)
\bibitem{b4} X. An, J. Wu, W. Tong, P. Zhu, and Y. Chen, {\it{6G Network Architecture Vision}}, in 2021 Joint European Conference on Networks and Communications \& 6G Summit (EuCNC/6G Summit), 2021, pp. 592–597.
\bibitem{b5} World Wide Web Consortium (W3C), {\it{Decentralized Identifiers v1.0}}. [Online]. Available: https://www.w3.org/TR/did-core/ (accessed: July 19 2022).
\bibitem{b6} K. C. Toth and A. Anderson-Priddy, {\it{Self-Sovereign Digital Identity: A Paradigm Shift for Identity}}, IEEE Security \& Privacy, vol. 17, no. 3, pp. 17–27, 2019.
\bibitem{b7} V.-L. Nguyen, P.-C. Lin, B.-C. Cheng, R.-H. Hwang, and Y.-D. Lin, {\it{Security and Privacy for 6G: A Survey on Prospective Technologies and Challenges}}, IEEE Communications Surveys \& Tutorials, vol. 23, no. 4, pp. 2384-2428, 2021.
\bibitem{b8} J. Xu, K. Xue, H. Tian, J. Hong, D. S. L. Wei, and P. Hong, {\it{An Identity Management and Authentication Scheme Based on Redactable Blockchain for Mobile Networks}}, IEEE Transactions on Vehicular Technology, vol. 69, no. 6, pp. 6688-6698, 2020.
\bibitem{b9} Z. Haddad, M. M. Fouda, M. Mahmoud, and M. Abdallah, {\it{Blockchain-based Authentication for 5G Networks}}, in 2020 IEEE Int. Conf. on Informatics, IoT, and Enabling Technologies (ICIoT), 2020, pp. 189–194.
\bibitem{b10} S. Raju, S. Boddepalli, S. Gampa, Q. Yan, and J. S. Deogun, {\it{Identity management using blockchain for cognitive cellular networks}}, in 2017 IEEE Int. Conf. on Communications (ICC), 2017, pp. 1–6.
\bibitem{b11} G. Fedrecheski, J. M. Rabaey, L. C. P. Costa, P. C. Calcina Ccori, W. T. Pereira, and M. K. Zuffo, {\it{Self-Sovereign Identity for IoT environments: A Perspective}}, in 2020 Global Internet of Things Summit (GIoTS), 2020, pp. 1–6.
\bibitem{b12} World Wide Web Consortium (W3C), {\it{Verifiable Credentials Data Model v1.1.}} [Online]. Available: https://www.w3.org/TR/vc-data-model/ (accessed: July 19 2022).
\bibitem{b13} R. Ansey, J. Kempf, O. Berzin, C. Xi, and I. Sheikh, {\it{Gnomon: Decentralized Identifiers for Securing 5G Iot Device Registration and Software Update}} in 2019 IEEE Globecom Workshops (GC Workshops),
2019, pp. 1–6.
\bibitem{b14} X. Salleras and V. Daza, {\it{SANS: Self-Sovereign Authentication for Network Slices}}, Security and Communication Networks, vol. 2020, pp. 1–8, 2020.
\bibitem{b15} X. Fan, Q. Chai, L. Xu, and D. Guo, {\it{DIAM-IoT: A Decentralized Identity and Access Management Framework for Internet of Things}} in Proceedings of the 2nd ACM Int. Symposium on Blockchain and Secure Critical Infrastructure,, Eds. ACM, 2020, pp. 186–191.
\end{thebibliography}
\end{document}